\documentclass[preprint
 ,secnumarabic
,amssymb,amsmath,nobibnotes,aps,pre,showpacs,floatfix]{revtex4}
\usepackage{bm}

\begin{document}

\title{On the computation of the entropy in the microcanonical ensemble
for mean-field-like systems}

\author{Alessandro Campa}
\email{campa@iss.infn.it}
\affiliation{Complex Systems and Theoretical Physics Unit, Health and Technology Department,
Istituto Superiore di Sanit\`a, and INFN Roma1 - Gruppo Collegato Sanit\`a,
Viale Regina Elena 299, 00161 Roma, Italy}

\date{\today}

\begin{abstract}
Two recently proposed expressions for the computation of the entropy
in the microcanonical ensemble are compared, and their equivalence
is proved. These expressions are valid for a certain class of
statistical mechanics systems, that can be called mean-field-like
systems. Among these, this work considers only the systems with the most
usual hamiltonian structure, given by a kinetic energy term plus interaction
terms depending only on the configurational coordinates.
\end{abstract}


\maketitle

The microcanonical ensemble generally is not the chosen ensemble for the
computation of thermodynamic observables of statistical systems. Exact or
approximate analytical procedures are usually easier in the canonical
ensemble. However, a recent surge of interest for the statistical
properties of systems with ``unconventional'' thermodynamics has
made desirable to perform calculations in both ensembles. These systems
are characterized by the fact that they are small~\cite{gross}, and thus
surface effects cannot be neglected, or by the long range nature of the
interaction between their microscopic constituents~\cite{draw}. In both cases
additivity of the energy and of other thermodynamic functions does not hold.

The most relevant feature of long range systems is that very often, even in
the thermodynamic limit, canonical and microcanonical ensembles are not
equivalent, and this justifies the necessity to perform calculations in both
ensembles: contrary to the usual case, the system can have different values
of the macroscopic observables, if it is considered at fixed energy or at fixed
temperature. The issue of ensemble inequivalence, first discussed theoretically
by Hertel and Thirring~\cite{heth}, has later been thoroughly treated in the
mathematical physics literature (see Refs.~\cite{elht,kiesleb} and references
therein). The inequivalence has been put in evidence both in self-gravitating
systems~\cite{lynd2,chav1,chav2} and in other, magnetic-type,
systems~\cite{cost,barre,bbdr}, which are often easier to study.
A fundamental tool for the mathematical analysis of ensemble inequivalence has been
found in the theory of large deviations~\cite{ellbook}, as applied to statistical
mechanics (see Ref.~\cite{ellisa} and references therein): this application employs
the important concept of macrostates of the systems~\cite{elht,touch,ellisb}.

In simple terms, nonequivalence can be ascribed to the nonconcavities that are found
in the microcanonical entropy~\cite{elht}. Not only does this lead to different
results for the entropy, if computed in the microcanonical and in the canonical ensembles,
but it can also be the cause, in the microcanonical ensemble, of negative specific
heats~\cite{lynd1,padm,lynd2,chav1,chav2,barre,cost,guch}. It has also been found
that ensemble inequivalence is associated to the presence of first order phase
transitions in the canonical ensemble~\cite{chav1,barre}.

Through the use of large deviation techniques, it is possible to derive a general
expression for the microcanonical entropy of statistical mechanics
systems belonging to a certain class~\cite{elht,bbdr,ellisa}, that we can call
mean-field-like systems. This terminology is motivated by the observation that
in these systems the interactions can be thought of as giving
rise to a mean field acting on the particles, although this may not be easily
visible as in, e.g., some simple lattice magnetic models.
If one accepts to sacrifice a full mathematical rigour, the derivation of this
expression can be short~\cite{bbdr,caru}. Recently, another expression has
been proposed~\cite{leru}, starting from the form that the canonical partition
function assumes in mean-field-like systems; no obvious and self-evident relation
with the large deviation expression can be easily found. In this work our purpose is
to make a connection between these two different expressions. Our motivation is to show
the equivalence of the two approaches, and at the same time to give an operative
procedure for the derivation of the particular form of the canonical partition function
in mean-field-like systems, that in Ref.~\cite{leru} is used as a starting point, but is
not derived.

In a system with $N$ degrees of freedom and canonical coordinates $\{(q_i,p_i)\}$, we are
interested in the computation of the entropy per particle
in the thermodynamic limit:
\begin{equation}\label{entrgen}
s(\epsilon)=\lim_{N\rightarrow \infty} \frac{1}{N} \ln \int \left(\Pi_{i=1}^N {\rm d}
q_i{\rm d}p_i\right) \,
\delta \left( H(\{(q_i,p_i)\})-N\epsilon\right),
\end{equation}
where $H$ is the Hamiltonian and $\epsilon$ is the energy per particle (or energy density).
Let us first define the class of systems that are considered here. We suppose that the
Hamiltonian of our systems can be written in the form:
\begin{equation}\label{hamil}
H = N \sum_{k=1}^n g_k(M_k),
\end{equation}
with
\begin{equation}\label{largvar}
M_k = \frac{1}{N} \sum_{i=1}^N f_k(q_i,p_i),
\end{equation}
where $f_k$ and $g_k$ are smooth functions. It should be pointed out that the theory
of large deviations can be applied to systems defined by looser conditions on the
Hamiltonian; in particular, the form (\ref{hamil}) could also be the leading
term of an expansion in $N$, in which the relative weight of the remaining
terms vanishes in the thermodynamic limit; even the simple additive
structure displayed in (\ref{hamil}) is not necessary~\cite{elht}. However, for
simplicity here we restrict to systems in which Eq.~(\ref{hamil}) holds
exactly; moreover, we treat the case in which the $k=1$ term is the kinetic energy,
i.e., $f_1(q,p)=f_1(p)=p^2/2$ (with a suitable choice of the mass unit) and
$g_1(x)=x$, while the functions $f_k$ for $k\ge 2$ represent the interaction terms,
and depend only of the configurational coordinate $q$. Thus, we are considering
mean-field-like systems in which each particle is described by one degree of freedom.
Later it will be shown that the extension to more than a degree of freedom per
particle (e.g., cartesian coordinates of particles in ordinary threedimensional
space) is easily performed.

We now show our first expression, that we call the large deviation expression
of the entropy per particle. The cited literature can be consulted for its
derivation. One finds that:
\begin{equation}\label{entrlg1}
s(\epsilon)=\sup_{m_1,\dots,m_n,\sum g_k(m_k)=\epsilon} J(m_1,\dots,m_n),
\end{equation}
where
\begin{equation}\label{entrlg2}
J(m_1,\dots,m_n)= \inf_{z_1,\dots,z_n} L(z_1,\dots,z_n,m_1,\dots,m_n)
\end{equation}
and where
\begin{equation}\label{entrlg3}
L(z_1,\dots,z_n,m_1,\dots,m_n)= \ln \langle {\rm e}^{\sum_{k=1}^n z_kf_k}
\rangle_1 -\left(\sum_{k=1}^nz_km_k\right);
\end{equation}
here $\langle \cdot \rangle_1$ is the one particle phase space (unnormalized)
average defined by:
\begin{equation}\label{onephsp}
\langle \cdot \rangle_1 = \int {\rm d}q{\rm d}p \, (\cdot).
\end{equation}
Of course, depending on the form of the functions $f_k$, there can be restrictions
on the sign of the $z_k$s, in order to have convergence of the one particle
phase space average in the right hand side of Eq.~(\ref{entrlg3}).

The second expression is obtained, as already mentioned, in Ref.~\cite{leru}, and
it is derived under the hypothesis that
the canonical partition function of the system can be written in the form:
\begin{eqnarray}\label{canmf}
Z(\beta)&\equiv&\int \left(\Pi_{i=1}^N {\rm d}q_i{\rm d}p_i\right)
\, {\rm exp}\left[-\beta H(\{(q_i,p_i)\})\right] \\
=&& C \int {\rm d}y_1\dots{\rm d}y_r \, {\rm exp}\left[N\phi(\beta,y_1,\dots,y_r)\right],
\nonumber
\end{eqnarray}
where the factor $C$ has at most a power law dependence on $N$, so that it does not
contribute, in the thermodynamic limit, to the free energy per particle $f(\beta)$,
given by:
\begin{eqnarray}\label{freepp}
-\beta f(\beta)&=&\lim_{N\rightarrow \infty} \frac{1}{N} \ln Z(\beta) \nonumber \\
&=&\sup_{y_1,\dots,y_r}\phi(\beta,y_1,\dots,y_r).
\end{eqnarray}
In the derivation of the microcanonical entropy, apart from the assumption of analiticity,
there is no other hypothesis about the form of the real function $\phi$, about the number
$r$ of ``order parameters'', and about the way in which $\phi$ has been obtained. Actually,
Ref.~\cite{leru} treats the case $r=1$, but the generalization to $r>1$ is
straightforward~\cite{acag}. It can then be shown that the microcanonical entropy per
particle is given, in the thermodynamic limit, by:
\begin{equation}\label{entrlr}
s(\epsilon)=\sup_{y_1,\dots,y_r}\left[ \inf_\beta \left(\beta \epsilon +
\phi(\beta,y_1,\dots,y_r)\right)\right].
\end{equation}
In Ref.~\cite{leru} the systems for which Eq.~(\ref{canmf}) holds are called
mean-field-like systems; thus we will call Eq.~(\ref{entrlr}) the mean-field-like
expression of the entropy per particle.
Our purpose here is to show that this last expression is equivalent to
Eqs.~(\ref{entrlg1}-\ref{entrlg3}).

First of all, let us specialize the expressions to our case, that is concerned with
the most usual situation in which in the Hamiltonian there is a kinetic energy term plus
terms depending only on the configurational coordinates and giving the interaction. Then,
in the one particle phase space average of Eq.~(\ref{entrlg3}), the integrals in
$q$ and $p$ separate; besides, the minimization problem
defined in Eq.~(\ref{entrlg2}) decouples in one concerning $z_1$, that is easily
solved, and one concerning $z_2,\dots,z_n$. One ends up with:
\begin{eqnarray}\label{entrblg1}
J(m_1,\dots,m_n)&&= \frac{1}{2}\ln (2\pi{\rm e}) + \frac{1}{2}\ln (2m_1)\nonumber \\
+\inf_{z_2,\dots,z_n}&& G(z_2,\dots,z_n,m_2,\dots,m_n),
\end{eqnarray}
where
\begin{eqnarray}\label{entrblg2}
&&G(z_2,\dots,z_n,m_2,\dots,m_n)\equiv G(\{z_k\},\{m_k\})\nonumber \\
&=& \ln \langle {\rm e}^{\sum_{k=2}^n z_kf_k}
\rangle_{1q} -\left(\sum_{k=2}^nz_km_k\right),
\end{eqnarray}
where now $\langle \cdot \rangle_{1q}$ is the analogous of Eq.~(\ref{onephsp}), but
with the integration only in $q$.
Let us denote with $\overline{z}_k$, $k=2,\dots,n$, the solution of the minimization
problem defined in the right hand side of Eq.~(\ref{entrblg1}); the $\overline{z}_k$s
are thus functions of $m_2,\dots,m_n$. Furthermore, since $g_1(x)=x$ in
Eq.~(\ref{hamil}), the maximation problem in Eq.~(\ref{entrlg1}) can be performed
by a free maximation over $m_2,\dots,m_n$ after putting:
\begin{equation}\label{m1def}
m_1=\epsilon - \sum_{k=2}^n g_k(m_k).
\end{equation}
Therefore, neglecting the constant term in the right hand side of Eq.~(\ref{entrblg1}),
we obtain:
\begin{eqnarray}\label{entrblg3}
s(\epsilon)=\sup_{m_2,\dots,m_n}&& \Bigg[ \frac{1}{2}\ln \Big(2\epsilon - 2\sum_{k=2}^n
g_k(m_k)\Big) \nonumber \\
 +&& G(\overline{z}_2,\dots,\overline{z}_n,m_2,\dots,m_n)\Bigg].
\end{eqnarray}
This is the large deviation expression of the thermodynamic limit of the entropy per
particle, for the class of systems considered here.

We can now show, starting from the definition of the canonical partition function,
that Eq.~(\ref{entrlr}) can be transformed in Eq.~(\ref{entrblg3}); moreover, we can
derive a definite form of $\phi$ in Eq.~(\ref{canmf}). In the computation of the canonical
partition function, it is possible to adopt a procedure that is similar to that employed
in the derivation of the large deviation expression of the entropy~\cite{caru}. In fact,
taking into account the form of the Hamiltonian, Eqs.~(\ref{hamil}) and (\ref{largvar}),
and specializing to our case, we can write:
\begin{eqnarray}\label{canlg1}
&&Z(\beta)=\left(\frac{2\pi}{\beta}\right)^{\frac{N}{2}}\int {\rm d}q_1\dots{\rm d}q_N
{\rm d}m_2\dots{\rm d}m_n \\
&&\big(\Pi_{k=2}^n N\delta(NM_k(\{q_i\}) -Nm_k)\big)
\exp\left[-N\beta \sum_{k=2}^n g(m_k)\right]. \nonumber
\end{eqnarray}
The phase space coordinates are now only in the $\delta$ functions. Using the inverse
Laplace transform for these functions we find that:
\begin{eqnarray}\label{canpar}
&&\int {\rm d}q_1\dots{\rm d}q_N
\big(\Pi_{k=2}^n N\delta(NM_k(\{q_i\}) -Nm_k)\big) \nonumber \\
&=&\int_{{\mathcal C}} {\rm d}z_2\dots{\rm d}z_n
\left(\frac{N}{2\pi{\rm i}}\right)^n \! \! \exp \left[ N
G(\{z_k\},\{m_k\})\right],
\end{eqnarray}
with the function $G$ defined in (\ref{entrblg2}), and where the integration
path for each of the variables $z_k$ is the imaginary axis, or a path obtained
by deformation of the imaginary axis without crossing singularities. 
We compute the integral in the right hand side of (\ref{canpar}), for
$N\rightarrow \infty$, with the saddle point method (full mathematical rigour
would require to prove the analiticity of the function $G$). To avoid unphysical
oscillatory behavior, the relevant stationary point must lie on the real
$z_k$ axes. It can then be inferred that this stationary point can be found
looking for the minimum of $G$ with respect to real values of $z_2,\dots,z_n$.
This can be deduced in the following way. Let us first define the weighted
one particle average $\langle \cdot \rangle_w$ by:
\begin{equation}\label{onewei}
\langle \cdot \rangle_w= \frac{\langle (\cdot){\rm e}^{\sum_{k=2}^n z_kf_k}
\rangle_{1q}}{\langle {\rm e}^{\sum_{k=2}^n z_kf_k}\rangle_{1q}}.
\end{equation}
The leading order variation $\Delta G$ around a stationary point of the function
$G$ is easily found, from Eq.~(\ref{entrblg2}), to be:
\begin{eqnarray}\label{varg}
\Delta G &=& \sum_{k=2}^n\sum_{l=2}^n\langle \left(f_k-\langle f_k \rangle_w\right)
\left(f_l-\langle f_l \rangle_w\right)\rangle_w {\rm d}z_k{\rm d}z_l \nonumber \\
&=&\langle \left( \sum_{k=2}^n \left(f_k-\langle f_k \rangle_w\right){\rm d}
z_k \right)^2 \rangle_w.
\end{eqnarray}
Restricting to real $z_k$s we have both that $G$ is a real function, and that
Eq.~(\ref{varg}) is a positive quantity, meaning that a stationary
point is a minimum. Thus the saddle point evaluation of the integral in the right
hand side of Eq.~(\ref{canpar}) is obtained traversing the relevant stationary point
along a line parallel to the imaginary axis for each $z_k$, so that the real part of
$G$ attains the maximum in the integration domain. At the end, keeping
only the exponential dependence on $N$, Eq.~(\ref{canpar}) becomes: 
\begin{eqnarray}\label{canpar2}
&&\int {\rm d}q_1\dots{\rm d}q_N
\big(\Pi_{k=2}^n N\delta(NM_k(\{q_i\}) -Nm_k)\big) \nonumber \\
&\propto& \exp \left[ N
G(\overline{z}_2,\dots,\overline{z}_n,m_2,\dots,m_n)\right].
\end{eqnarray}
This result can be inserted in Eq.~(\ref{canlg1}), and one therefore obtains an
expression like in Eq.~(\ref{canmf}), with $r=n-1$ and, after the change of
notation $y_1,\dots,y_r\rightarrow m_2,\dots,m_n$, with $\phi$ given by:
\begin{eqnarray}\label{phidev}
\phi(\beta,m_2,\dots,m_n)&=&\frac{1}{2}\ln \left(\frac{2\pi}{\beta}\right)
-\beta \sum_{k=2}^n g(m_k) \nonumber \\
&+& G(\overline{z}_2,\dots,\overline{z}_n,m_2,\dots,m_n).
\end{eqnarray}
At this point, with the very simple dependence on $\beta$ of this expression, it
is straightforward to perform the minimization problem in $\beta$ indicated in
Eq.~(\ref{entrlr}); we obtain:
\begin{equation}\label{betasol}
\frac{1}{\beta} = 2\epsilon - 2 \sum_{k=2}^n g(m_k).
\end{equation}
This is substituted in $\beta \epsilon + \phi(\beta,m_2,\dots,m_n)$, and then the
mean-field-like expression of the microcanonical entropy per particle,
Eq.~(\ref{entrlr}), becomes, again neglecting the same constant term as before,
exactly the same as the large deviation expression, Eq.~(\ref{entrblg3}).

The extension to systems in which the particles move in a D-dimensional space
is straightforward. In this case the canonical coordinates $\{q_i,p_i\}$ in
Eqs.~(\ref{entrgen}), (\ref{largvar}), and (\ref{canmf}) are to be intended as
D-dimensional vectors $\{\mathbf{q}_i,\mathbf{p}_i\}$. Analogous replacements
have to be done for the configurational coordinates $q_i$ in Eq.~(\ref{canpar}),
in the one particle phase space average $\langle \cdot \rangle_1$ (see
Eq.~(\ref{onephsp})), and in its configurational restriction
$\langle \cdot \rangle_{1q}$. With $\epsilon$ we still denote the energy per
particle (not per degree of freedom). At the end, the common result for the 
large deviation and the mean-field-like expressions of the entropy per particle,
Eq.~(\ref{entrblg3}), is simply generalized, neglecting constant terms, to:
\begin{eqnarray}\label{entrblg4}
s(\epsilon)=\sup_{m_2,\dots,m_n}&& \Bigg[ \frac{D}{2}\ln \Big(2\epsilon - 2\sum_{k=2}^n
g_k(m_k)\Big) \nonumber \\
 +&& G(\overline{z}_2,\dots,\overline{z}_n,m_2,\dots,m_n)\Bigg];
\end{eqnarray}

We end with the following observation. The form obtained for the function $\phi$ in
Eq.~(\ref{phidev}) is useful for our proof of equivalence of the two approaches to
the computation of the microcanonical entropy per particle, but by no means it is
unique. For example, with simple quadratic functions $g_k$ in Eq.~(\ref{hamil}),
it is common to employ the Hubbard-Stratonovich transformation~\cite{hubb} to
linearize the dependence of the Hamiltonian on the phase space variables. One then
arrives, for the canonical partition function, to an expression as in the righmost
hand side of Eq.~(\ref{canmf}), with the functional form of $\phi$, and the physical
meaning of the ``order parameters'' $m_k$, that can be different from that in
Eq.~(\ref{phidev}). Depending on the concrete cases, the function $\phi$ obtained 
may result in an expression that is easier to analyze numerically than
Eq.~(\ref{phidev}).

\end{document}